\documentclass[a4paper,10pt,twocolumn]{article}
\usepackage[a4paper, margin=2cm]{geometry}

\usepackage[utf8]{inputenc}
\usepackage[T1]{fontenc}

\usepackage{fourier}

\usepackage[authoryear]{natbib}
\usepackage[fleqn]{amsmath}
\allowdisplaybreaks
\usepackage{amsfonts}
\usepackage{bm}
\usepackage{mathrsfs}
\usepackage{booktabs}
\usepackage{graphicx}
\usepackage{url}
\usepackage{xcolor}
\usepackage[only,llbracket,rrbracket]{stmaryrd}

\usepackage{amsthm}
\theoremstyle{remark}

\usepackage[english]{babel}
\usepackage[hidelinks]{hyperref}

\usepackage{doi}

\usepackage{ulem} 

\title{Singular travelling waves in soft viscoelastic solids of rate type}
\author{Harold Berjamin \textsuperscript{a}, Michel Destrade \textsuperscript{a,b}, Giuseppe Saccomandi \textsuperscript{c,a} \\ ~ \\
\emph{\textsuperscript{a}School of Mathematical and Statistical Sciences, University of Galway,}\\ \emph{University Road, Galway, Republic of Ireland} \\
\emph{\textsuperscript{b}Key Laboratory of Soft Machines and Smart Devices }\\ \emph{of Zhejiang Province and Department of Engineering Mechanics, Zhejiang University,}\\ \emph{Hangzhou 310027, People's Republic of China} \\
\emph{\textsuperscript{c}Dipartimento di Ingegneria, Università degli studi di Perugia,}\\ \emph{Via G. Duranti, Perugia 06125, Italy} }

\date{}

\begin{document}

\twocolumn[	
\begin{@twocolumnfalse}

	\maketitle
	
	\begin{abstract}
		\noindent
		We consider shear wave propagation in soft viscoelastic solids of rate type. Based on objective stress rates, the constitutive model accounts for finite strain, incompressibility, as well as stress- and strain-rate viscoelasticity. The theory generalises the standard linear solid model to three-dimensional volume-preserving motions of large amplitude in a physically-consistent way. The nonlinear equations governing shear motion take the form of a one-dimensional hyperbolic system with relaxation. For specific objective rates of Cauchy stress (lower- and upper-convected derivatives), we study the propagation of acceleration waves and shock waves. Then we show that both smooth and discontinuous travelling wave solutions can be obtained analytically. We observe that the amplitude and velocity of steady shocks are very sensitive to variations of the stress relaxation time. Furthermore, the existence of steady shocks is conditional. Extension of these results to the case of multiple relaxation mechanisms and of the Jaumann stress rate is attempted. The analysis of simple shearing motions is more involved in these cases. \\
		~ \\
		\emph{Keywords:~} Travelling waves, Shock waves, Nonlinear viscoelasticity, Soft solids, Hyperbolic systems
	\end{abstract}

	\vspace{1em}

\end{@twocolumnfalse}
]

\section{Introduction}\label{sec:Intro}

Soft solids are materials that can be easily deformed through the application of external forces. In Physics and Engineering, this term is commonly used for highly deformable materials such as elastomers \citep{haupt02}, bitumen \citep{filograna09}, dough \citep{phanthien97}, as well as soft biological tissues \citep{berjamin22a}. Due to their low stiffness, soft solids are routinely subjected to large deformations and dynamic motions of arbitrary frequency.

The mechanical modelling of soft solids has been approached both by the fluid mechanics and solid mechanics communities. In both cases, incompressibility is commonly assumed, thus restricting the motion to volume-preserving deformations (e.g., simple shear or pure torsion). Over the years, various incompressible viscoelasticity theories have also been proposed to describe the dynamic behaviour of soft solids, including the modelling of dissipation. Plastic deformations are usually neglected in soft solids for most applications given their large yield strength.

Due to the occurrence of large deformations, related continuum theories must be formulated within the finite strain theory \citep{holzapfel00}, in general. The simplest model consists of a Newtonian viscous stress added to an elastic stress contribution \citep{berjamin23}, which results in a nonlinear three-dimensional version of the classical Kelvin--Voigt viscoelasticity theory. Unfortunately, such a model is valid in the low frequency range only, and it is not able to reproduce relaxation of stress \citep{banks11,carcione15}.

To overcome these issues, several versions of stress- and strain-rate incompressible viscoelasticity are found in the literature. Generalising \citet{cormack18} to three-dimensional motions, \citet{saccomandi21} introduce a three-dimensional Maxwell theory similar to the models presented by \citet{haupt02}. In all these cases, the constitutive law is augmented by an evolution equation for the viscous stress that reduces to the Kelvin--Voigt model in a given limit. As pointed out by \citet{morro20}, the process leading to stress rate viscoelasticity theories is very similar to the derivation of the Maxwell--Cattaneo--Vernotte hyperbolic theory of heat conduction.

Another popular approach, known as quasi-linear viscoelasticity (QLV), does not include any explicit constitutive assumption for the description of the viscous stresses. Instead, the viscoelastic stress response is directly deduced from the inviscid elastic response by means of a hereditary integral (Boltzmann superposition). The latter can be converted into linear differential equations when the hereditary integral is based on a Prony series \citep{berjamin22a}. In this case, connections with the Simo model can be established, see the discussions in \citet{yagimli23}.

One feature of the dynamic response of nonlinear viscoelastic materials is the existence of travelling wave solutions in shear, aka steady progressive waves. These waves have the particularity of keeping an invariant wave profile throughout the motion which occurs at a suitable constant speed. Similarly to other solitary waves, those permanent waveforms result from the interaction between nonlinearity and dispersion (here of dissipative nature).

In the Kelvin--Voigt case, smooth travelling wave solutions can be derived exactly, making this one of the few examples of a purely analytical solution in nonlinear viscoelasticity \citep{berjamin23}. Implicit analytical expressions can also be obtained for a specific history-dependent material model \citep{pucci15}. In the QLV case, the study of solitary waves in shear is less straightforward. Nevertheless, smooth travelling wave solutions can still be obtained numerically \citep{depascalis19}.

For the model considered by \citet{pucci15}, it was reported that the wave profile can become multi-valued under certain conditions, suggesting the existence of weak discontinuous solutions in shear. Discontinuous travelling waves were also described by \citet{cormack18} based on an approximate one-dimensional theory, which involves a generalised Burgers-type equation that governs shearing motions \citep{berjamin23}.

The existence of \emph{steady shock waves} in compression is extensively discussed in the review by \citet{schuler73}, which covers various theoretical and experimental results related to the study of such waves. In particular, it was shown that the theory of steady shocks correlates well with one-dimensional experimental results obtained in a specific configuration. Furthermore, similar discontinuous wave profiles were obtained in torsion by \citet{sugimoto84} based on a thin rod approximation \citep{sugimoto84b}.

In relation with the above literature, we consider shear wave propagation in materials of rate type governed by the three-dimensional theory of \citet{saccomandi21}, whose properties are summarised in Section~\ref{sec:GovEq}. 
We rewrite the equations of motion as a one-dimensional hyperbolic system of conservation laws with relaxation, in the particular case of the lower- and upper-convected objective stress rates \citep{haupt02}. The propagation of shear acceleration waves and shear shock waves is studied in Section~\ref{sec:Shear}, and travelling wave solutions are detailed in Section~\ref{sec:Steady}, including smooth kinks and steady shocks. Finally, we reconsider these derivations in the case of multiple relaxation mechanisms (Section~\ref{sec:Multiple}) and in the case of the Jaumann stress rate (Section~\ref{sec:Jaumann}).

In summary, we show that steady shear shock waves can propagate in soft solids of rate type, by exploiting the simplifications provided by the lower- and upper-convected stress rates \citep{saccomandi21}. This way, fully analytical travelling wave solutions are obtained, including discontinuous solutions. These results are obtained using a three-dimensional theory specialised to simple shear motions, but without relying on any further approximation, see \citet{berjamin23} for a presentation of common approximations used in nonlinear acoustics. Moreover, we provide elements indicating that a similar behaviour might be obtained for multiple relaxation mechanisms and for the Jaumann stress rate, even though the complete mathematical derivations are less simple in these cases.

\section{Basic equations}\label{sec:GovEq}

\subsection{Constitutive model}\label{subsec:Constit}

Let us introduce the deformation gradient tensor $\bm{F} = \partial{\bm x}/\partial{\bm X}$, which is the gradient of the current position $\bm x$ of a particle with respect to its reference position $\bm X$. We introduce also the displacement field $\bm{u} = \bm{x}-\bm{X}$.
In incompressible materials, volume change is not allowed, so that $\bm F$ is unimodular ($\det \bm{F} \equiv 1$). Moreover, the mass density $\rho > 0$ is constant. Here, the spatial and temporal coordinates are omitted, but we implicitly assume $\bm{F} = \bm{F}(\bm{X},t)$, etc.

We assume that the Cauchy stress tensor $\bm{T}$ may be decomposed additively as
\begin{equation}
	\bm{T} = -p\bm{I} + {\bm T}^\text{E}, \qquad {\bm T}^\text{E} = \bm{T}^\text{e} + \bm{T}^\text{v},
	\label{Constitutive}
\end{equation}
where $\bm{T}^\text{e}$ is an elastic contribution, $\bm{T}^\text{v}$ is a viscous one, and their sum ${\bm T}^\text{E}$ is the extra stress. The first term of \eqref{Constitutive}\textsubscript{a} involves the identity tensor $\bm I$ and a scalar Lagrange multiplier $p = p(\bm{X}, t)$ accounting for the incompressibility constraint. The undetermined stress $-p\bm{I}$ is presented in Section~6.3 of \citet{holzapfel00} in the elastic case; the same principle holds in the viscoelastic case \citep{haupt02}.

We assume that the elastic stress contribution is of Mooney--Rivlin--Yeoh type:
\begin{equation}
	\bm{T}^\text{e} = 2 C_1 \left(1 + \beta\, (\text{tr}\bm{B}-3)\right) \bm{B} - 2C_2\bm{B}^{-1} , 
	\label{ConstitElast}
\end{equation}
where $\bm{B} = \bm{F}\bm{F}^\top$ is the left Cauchy--Green strain tensor. Here, $C_1$, $C_2$ are the Mooney parameters and $\beta$ is a coefficient of nonlinearity. The infinitesimal shear modulus equals $\mu = 2\, (C_1+C_2) > 0$. The above constitutive law has been used in relation with the modelling of brain tissue \citep{berjamin22a}, but other choices of elastic response are possible.

The evolution of the viscous stress is governed by the Maxwell-type differential equation \citep{saccomandi21}
\begin{equation}
	\bm{T}^\text{v} + \tau\, \mathscr{D}(\bm{T}^\text{v}) =  2\eta \bm{D} ,
	\label{ConstitVisc}
\end{equation}
where $\eta > 0$ is the shear viscosity and $\tau>0$ is a characteristic time. 
Here, the strain rate tensor $\bm{D} = \frac12(\bm{L} + \bm{L}^\top)$ is the symmetric part of the Eulerian velocity gradient $\bm{L} = \dot{\bm F}\bm{F}^{-1}$, which involves the material time-derivative $\dot{\bm F}$ of $\bm{F}$. The tensor $\mathscr{D}(\bm{T}^\text{v})$ is a rate of viscous stress to be specified subsequently. By virtue of incompressibility, we note that the tensors $\bm L$, $\bm D$ are trace-free.

In Eq.~\eqref{ConstitVisc}, the stress tensor $\bm{T}^\text{v}$ is \emph{objective} or \emph{frame-indifferent}, meaning that it does not depend on the motion of the observer \citep{holzapfel00}. Furthermore, while $\dot{\bm F}$ and $\bm{L}$ are not objective, the strain rate tensor $\bm D$ is objective. Thus, to ensure the consistency of Eq.~\eqref{ConstitVisc} with respect to objectivity, the stress rate $\mathscr{D}(\bm{T}^\text{v})$ therein needs to be objective as well. In general, $\mathscr{D}(\bm{T}^\text{v})$ should not be replaced by $\dot{\bm T}^\text{v}$, which is not objective.

Suitable choices for $\mathscr{D}(\bm{T}^\text{v})$ are not unique. Here, we consider an objective rate defined by
\begin{equation}
	\mathscr{D}(\bm{T}) = \dot{\bm T}
	+ \tfrac{\phi-1}{2} \left(\bm{L}\bm{T} + \bm{T}\bm{L}^\top\right)
	+ \tfrac{\phi+1}{2} \left(\bm{L}^\top\bm{T} + \bm{T}\bm{L}\right)
	\label{ObjectiveDer}
\end{equation}
where $\phi \in \lbrace -1, 0, 1\rbrace$. The expression \eqref{ObjectiveDer} combines the `upper-convected' Oldroyd rate for $\phi=-1$ (also equivalent to the Truesdell rate in the incompressible case); the `co-rotational' Jaumann--Zaremba rate for $\phi=0$; and the `lower-convected' Cotter--Rivlin rate for $\phi=1$, see definitions in the monograph by \cite{holzapfel00}. This way, one ensures that the rate equation \eqref{ConstitVisc} is frame-indifferent.

Let us compute the objective rate \eqref{ObjectiveDer} of the extra stress ${\bm T}^\text{E}$, as defined in \eqref{Constitutive}\textsubscript{b}. Using the rate equation \eqref{ConstitVisc}, we arrive at
\begin{equation}
	\bm{T}^\text{E} + \tau\, \mathscr{D}(\bm{T}^\text{E}) = \bm{T}^\text{e} + 2\eta\bm{D} + \tau\, \mathscr{D}(\bm{T}^\text{e}) .
	\label{ConstitutiveExtra}
\end{equation}
This equivalent formulation of the above constitutive model allows direct comparison with other approaches found in the literature. In particular, \citet{filograna09} do not include the term $\tau\, \mathscr{D}(\bm{T}^\text{e})$ in the right-hand side of \eqref{ConstitutiveExtra}, see Eq.~\eqref{Filograna}.

The rate equation \eqref{ConstitVisc}-\eqref{ObjectiveDer} with $\phi = -1$ was also proposed by \citet{phanthien97} to model the viscoelastic behaviour of dough. For the cases $\phi = \pm 1$, thermodynamic consistency is addressed by \citet{haupt02} and, therefore, the related stress-rate theories are physically admissible. Thermodynamic consistency is proved by \citet{morro20} in the case $\phi=0$. Essentially, these results mean that there is no need for additional restrictions of the material parameters to avoid physical inconsistencies for all $\phi$, which contrasts with the observations made by \citet{farina22} in tensile creep. We discuss potential issues with the latter results in the Appendix~\ref{sec:Appendix} where the tensile creep problem is carefully reconsidered.

Up to a suitable redefinition of the arbitrary Lagrange multiplier $p$ accounting for incompressibility, we observe that the constitutive law \eqref{Constitutive}-\eqref{ConstitVisc} can be rewritten as
\begin{equation}
	\begin{aligned}
		&\bm{T} = -p\bm{I} + {\bm T}^\text{e}_\text{d} + {\bm T}^\text{v}_\text{d} , \\
		&{\bm T}^\text{v}_\text{d} + \tau\, \mathscr{D}(\bm{T}^\text{v})_\text{d} =  2\eta \bm{D} ,
	\end{aligned}
	\label{ConstitDev}
\end{equation}
where $(\bullet)_\text{d} = (\bullet) - \frac13 \text{tr}(\bullet) \bm{I}$ defines the deviator operator. In passing, it is worth pointing out that the deviator of an objective second-order tensor is an objective quantity as well. Thus, the evolution equation \eqref{ConstitDev}\textsubscript{b} is naturally frame-indifferent and trace-free.

This evolution equation is slightly different to the one in \citet{liapidevskii11}, where the deviatoric stress rate $\mathscr{D}(\bm{T}^\text{v})_\text{d}$ is replaced by
\begin{equation}
	\begin{aligned}
		\mathscr{D}(\bm{T}^\text{v}_\text{d}) &= \mathscr{D}(\bm{T}^\text{v})_\text{d} + \tfrac13 \text{tr}(\mathscr{D}(\bm{T}^\text{v}))\bm{I} - \tfrac13 \mathscr{D}(\text{tr}(\bm{T}^\text{v})\bm{I}) \\
		&= \mathscr{D}(\bm{T}^\text{v})_\text{d} + \tfrac23 \phi \left( \text{tr}({\bm T}^\text{v}\bm{D}) \bm{I} -  \text{tr}(\bm{T}^\text{v}) \bm{D} \right) .
	\end{aligned}
	\label{Pukh}
\end{equation}
For this modified model, \citet{pukhnachev10} selects the Jaumann derivative $\phi = 0$ in order to keep the evolution of deviatoric viscous stresses trace-free. Furthermore, thermodynamic consistency is proved in this special case where the relationship $\mathscr{D}(\bm{T}^\text{v}_\text{d}) = \mathscr{D}(\bm{T}^\text{v})_\text{d}$ is satisfied. Therefore, this result is coherent with that obtained by \citet{morro20}.

Eq. \eqref{ConstitVisc} can be integrated in time as \citep{goddard66}
\begin{equation}
	\bm{T}^\text{v} = \frac{2\eta}{\tau} \int_{-\infty}^t \text{e}^{-(t-s)/\tau} \bm{\Theta}_{t|s} \bm{D}(s) \bm{\Theta}_{t|s}^{\top} \text d s ,
	\label{Memory}
\end{equation}
where $\bm{\Theta}_{t|s}$ is defined as the fundamental matrix of the initial-value problem
\begin{equation}
	\dot{\bm \Theta}_{t|s} = -\tfrac{\phi-1}{2} \bm{L} {\bm \Theta}_{t|s} - \tfrac{\phi+1}{2} \bm{L}^\top {\bm \Theta}_{t|s} , \qquad
	{\bm \Theta}_{s|s} = \bm{I}.
	\label{Matrizant}
\end{equation}
This result can be verified by evaluation of $\dot{\bm T}^\text{v}$ from \eqref{Memory} based on the Leibniz integral rule. Again, it should be understood that the Lagrangian coordinates $\bm X$ are used here.

We note that Eqs.~\eqref{Memory}-\eqref{Matrizant} correspond to the \citet{johnson77} fluid model, whose parameter `$a$' equals $-\phi$, see Eqs.~(2.18)-(2.25) therein. With this connection in mind, the authors of the above study remark that $\phi=-1$ agrees with the affine molecular model by \citet{lodge74}. For other related models, the reader is referred to Chapter~4 of \citet{macosko94}.

From the integral expression \eqref{Memory} of the viscous stress, we observe that the tensor ${\bm T}^\text{v}$ is a memory variable whose current value at time $t$ depends on the motion's history. Using the relationship $\bm{L} = \dot{\bm F}\bm{F}^{-1}$, direct integration of \eqref{Matrizant} yields
\begin{equation}
	{\bm \Theta}_{t|s} = -\tfrac{\phi-1}{2} {\bm F}_{t|s} + \tfrac{\phi+1}{2} {\bm F}_{t|s}^{-\top} , \qquad
	\phi=\pm 1,
\end{equation}
where ${\bm F}_{t|s} = \bm{F}(t)\bm{F}^{-1}(s)$ is the relative deformation gradient from the configuration at the intermediate time $s$ to the configuration at the current time $t$. For $\phi = \pm 1$, the above expression provides useful simplifications of \eqref{Memory}, see also \cite{haupt02}. Unfortunately, no such formula is known in the case $\phi = 0$, to the authors' present knowledge.

\subsection{Equations of motion}\label{subsec:Eq}

The Lagrangian equations of motion in strong form read \citep{holzapfel00}
\begin{equation}
	\dot{\bm F} = \nabla \bm{v}, \qquad \rho \dot{\bm v} = \nabla\cdot\bm{P},
	\label{EqMot}
\end{equation}
where $\bm{v} = \dot{\bm x}$ is the velocity field, $\bm{P} = \bm{T}\bm{F}^{-\top}$ is the first Piola--Kirchhoff stress tensor, and $\nabla$ is the gradient operator in the material description (i.e., partial differentiation is performed with respect to the reference position $\bm X$). For the purpose of potential finite element implementations, the reader is referred to Chapter~8 of \citet{holzapfel00} where variational principles accounting for incompressibility are presented.

According to Eq.~\eqref{Constitutive}, the first Piola--Kirchhoff stress tensor can be decomposed as $\bm{P} = -p\bm{F}^{-\top} + \bm{P}^\text{E}$, where $-p\bm{F}^{-\top}$ is a constitutively undetermined reaction stress that accounts for incompressibility, and $\bm{P}^\text{E} = \bm{T}^\text{E}\bm{F}^{-\top}$ is the extra stress contribution deduced from the constitutive law \eqref{ConstitElast}-\eqref{ConstitVisc}. The divergence operator in Eq.~\eqref{EqMot} satisfies $[\nabla\cdot\bm{P}]_i=P_{ij,j}$ where indices after the comma denote spatial differentiation, and summation over repeated indices is performed (Einstein notation).

\paragraph{Infinitesimal-strain limit.} Let us linearise the equations of motion \eqref{EqMot} with respect to the displacement gradient tensor $\nabla{\bm u} = \bm{F}-\bm{I}$ about a stress-free undeformed state. The Piola--Kirchhoff stresses satisfy $\bm{P} \simeq \bm{T}$, and the elastic stress contribution reduces to
\begin{equation}
	\bm{T}^\text{e} = 2\mu\bm{\varepsilon}, \qquad \bm{\varepsilon} = \tfrac12 (\nabla{\bm u}+\nabla^\top{\bm u}) ,
	\label{LinElast}
\end{equation}
where $\bm{\varepsilon}$ is the infinitesimal strain tensor. Given that the stress rates satisfy $\mathscr{D}(\bm{T}) \simeq \dot{\bm T}$ in the linear limit, the evolution equation \eqref{ConstitVisc} for the viscous stress becomes 
\begin{equation}
	\bm{T}^\text{v} + \tau \dot{\bm T}^\text{v} = 2\eta\dot{\bm \varepsilon} ,
	\label{LinVisc}
\end{equation}
for all $\phi$.

Next, we compute ${\bm T} + \tau \dot{\bm T}$ using the decomposition \eqref{Constitutive} and the linearised equations \eqref{LinElast}-\eqref{LinVisc}. Following the application of the deviator operator $(\bullet)_\text{d}$ introduced in Eq.~\eqref{ConstitDev}, we recover the tensorial \emph{Standard Linear Solid} (SLS) differential equation for the deviatoric stress,
\begin{equation}
	{\bm T}_\text{d} + \tau \dot{\bm T}_\text{d} = 2\mu \left(\bm{\varepsilon}_\text{d} + \tau_\varepsilon \dot{\bm \varepsilon}_\text{d} \right), \qquad \tau_\varepsilon = \tau + \frac{\eta}{\mu},
	\label{SLS}
\end{equation}
where we have used the linearised incompressibility condition $\operatorname{tr}\, \bm{\varepsilon} = 0$.

Let us postulate the harmonic plane-wave form $\bm{u} \propto \text{e}^{\text i (\omega t - \kappa \bm{n}\cdot\bm{X})}$ where $\omega$ is the angular frequency, $\kappa$ is the wave number, and $\bm n$ is a unit vector. Injecting this Ansatz in the incompressibility condition $\operatorname{tr}\, \bm{\varepsilon} = 0$ gives us the orthogonality condition $\bm{u}\cdot \bm{n} = 0$. Next, the linearised equations of motion lead to the classical SLS dispersion relationship \citep{carcione15}
\begin{equation}
	\rho \frac{\omega^2}{\kappa^2} = \mu \frac{1+\text{i}\omega\tau_\varepsilon}{1+\text{i}\omega\tau} = \mu_\infty \left(1 - \frac{g}{1+\text i\omega\tau}\right),
	\label{DispersionSLS}
\end{equation}
with $\mu_\infty = \mu \tau_\varepsilon/\tau$ and $g = 1-\tau/\tau_\varepsilon$.

Some related properties follow straightforwardly. In fact, for a real angular frequency $\omega > 0$, the wave dissipation factor deduced from \eqref{DispersionSLS} equals \citep{carcione15}
\begin{equation}
	-\frac{\mathfrak{Im}(\kappa^2)}{\mathfrak{Re}(\kappa^2)} = \frac{\omega\, (\tau_\varepsilon - \tau)}{1 + \omega^2 \tau_\varepsilon \tau} ,
	\label{QSLS}
\end{equation}
which is maximum at the frequency $\omega_c = 1/\sqrt{\tau_\varepsilon \tau}$. The dissipation factor \eqref{QSLS} should remain positive. Here, we observe that the condition $\tau_\varepsilon > \tau$ ensuring dissipative behaviour is naturally satisfied given Eq.~\eqref{SLS}. Hence, we have $1/\tau_\varepsilon < \omega_c < 1/\tau$.

The present model has two elastic limits corresponding to low frequency, $\omega \ll 1/\tau_\varepsilon$ (or long time), and high frequency, $\omega \gg 1/\tau$ (or short time). Using the above expression of the dynamic modulus \eqref{DispersionSLS}, we observe that the low-frequency equilibrium shear modulus equals $\mu$, whereas the high-frequency instantaneous shear modulus equals $\mu_\infty$.

These properties can be inferred directly from the constitutive law. In fact, for extremely slow motions (i.e., with small velocity gradients and small stress rates), the evolution equation \eqref{ConstitVisc} yields ${\bm T}^\text{v} \simeq {\bm 0}$. Thus, the extra stress satisfies ${\bm T}^\text{E} \simeq {\bm T}^\text{e}$, whose effective shear modulus is $\mu$.

For very fast motions (i.e., with large velocity gradients and large stress rates), the evolution equation \eqref{ConstitVisc} becomes $\mathscr{D}({\bm T}^\text{v}) \simeq  2\eta \bm{D}/\tau$. Upon time integration, we therefore find ${\bm T}^\text{v} = 2\eta/\tau \int_{-\infty}^t {\bm\Theta}_{t|s} {\bm D}(s) {\bm\Theta}_{t|s}^\top \text d s$, which becomes ${\bm T}^\text{E} \simeq 2\mu_\infty\bm{\varepsilon}$ in the infinitesimal-strain limit. The effective shear modulus is $\mu_\infty$ in this limit.

\section{Plane shear waves}\label{sec:Shear}

\subsection{Governing equations}\label{subsec:ShearEq}

Given that the material is isotropic, let us consider a plane shear wave $\bm{u} = u(X,t) \bm{e}_2$ propagating along $X$ and polarised along $Y$, the first and second components of the Lagrangian position vector in orthonormal Cartesian coordinates. Hence, the components of the deformation gradient tensor are
\begin{equation}
	\bm{F} = \begin{bmatrix}
		1 & 0 & 0 \\
		\gamma & 1 & 0 \\
		0 & 0 & 1
	\end{bmatrix} ,
	\label{SimpleShear}
\end{equation}
where $\gamma = \partial_X u$ is the shear strain. For the present simple shear deformation, we note that $x=X$. Furthermore, since $\det \bm{F}\equiv 1$, the restriction to volume-preserving motions is naturally satisfied.

Now, we write the equations of motion \eqref{EqMot} for the simple shear deformation \eqref{SimpleShear}. We introduce the shear velocity $v = \partial_t u$ and the quantities
\begin{equation}
	%\begin{aligned}
	\mu r = T^\text{v}_{21}, \qquad
	\mu s = -\tfrac{\phi-1}{2}T^\text{v}_{11}-\tfrac{\phi+1}{2}T^\text{v}_{22}
	%\end{aligned}
	\label{MemVars}
\end{equation}
deduced from the components of the viscous stress, which correspond to rescaled viscous shear stresses or viscous compression stresses, respectively. Thus, we arrive at the first-order system of partial differential equations
\begin{equation}
	\left\lbrace
	\begin{aligned}
		\gamma_t &= v_x ,\\
		\rho v_t &= \sigma_x ,\\
		\tau r_t &= \left(\eta/\mu + \tau s\right) v_x - r ,\\
		\tau s_t &= (\phi^2-1)\tau r v_x - s ,
	\end{aligned}
	\right.
	\label{SystHyp}
\end{equation}
where indices indicate partial differentiation.
The shear stress $\sigma = P_{21} = T_{21}$ deduced from the constitutive law takes the form
\begin{equation}
	\sigma = \mu \left( \gamma + b\gamma^3 + r\right) ,
	\label{ShearStress}
\end{equation}
where $b = \beta C_1/\mu$ is a parameter of nonlinearity.

Looking at the system \eqref{SystHyp}, it appears that the case $\phi=\pm 1$ offers useful simplifications. In fact, we then have $s = s_0(x)\, \text{e}^{-t/\tau}$ where $s_0$ is an arbitrary function to be determined. Thus, these simplifications produce $s \equiv 0$ if the material is initially at rest, which will be true in most of the cases considered hereinafter. Then, the system \eqref{SystHyp} reduces to a first-order system of balance laws with one memory variable describing the history of viscous shear stresses, $r$. This system is very similar to the one-dimensional model proposed by \citet{cormack18}. It is also similar to the quasi-linear viscoelastic system (QLV) of \citet{berjamin22a,depascalis19}, although it is even simpler. In fact, it involves only three partial differential equations instead of four.

\paragraph{Dimensionless form.}

Let us introduce the reference duration $t_\text{r} = \eta/\mu$ and distance $x_\text{r} = t_\text{r} \sqrt{\mu/\rho}$, where $\mu$ is the shear modulus. Thus, the scaled coordinates $\bar{t} = t/t_\text{r}$ and $\bar{x} = x/x_\text{r}$ are dimensionless. The scaling procedure is then achieved by introducing the dimensionless quantities
\begin{equation}
	\bar\gamma = \gamma \sqrt{b}, \quad \bar v = \frac{v\sqrt{b}}{x_\text{r}/t_\text{r}} , \quad \bar r = r \sqrt{b} , \quad \bar\tau = \frac{\tau}{t_\text{r}} .
	\label{SystHypAdim}
\end{equation}
Thus, from Eqs.~\eqref{SystHyp}-\eqref{ShearStress} with $\phi = \pm 1$, we arrive at
\begin{equation}
	\left\lbrace
	\begin{aligned}
		\gamma_{t} &= v_{x} ,\\
		v_{t} &= (\gamma + \gamma^3 + r)_{x} , \\
		\tau r_{t} &= v_{x} - r ,
	\end{aligned}
	\right.
	\label{SystAdim}
\end{equation}
in which all the variables should have an overbar (the latter is omitted here for the sake of simplicity).

The above dimensionless form obtained for $\phi = \pm 1$ will be studied hereinafter. For this purpose, we introduce the conservative system of balance laws with relaxation
\begin{equation}
	{\bf q}_t + {\bf f}({\bf q})_x = {\bf R} {\bf q} ,
	\label{SystVect}
\end{equation}
where ${\bf q} = [\gamma,v,r]^\top$,
\begin{equation}
	{\bf f}({\bf q}) = -\begin{bmatrix}
		v\\
		\gamma + \gamma^3 + r\\
		v/\tau
	\end{bmatrix} , \quad
	{\bf R} = -\begin{bmatrix}
		0 & 0 & 0\\
		0 & 0 & 0\\
		0 & 0 & 1/\tau
	\end{bmatrix} .
	\label{SystVar}
\end{equation}
For this system, the eigenvalues of the diagonalisable Jacobian matrix
\begin{equation}
	{\bf A}({\bf q}) = -\begin{bmatrix}
		0 & 1 & 0\\
		1 + 3\gamma^2 & 0 & 1\\
		0 & 1/\tau & 0 
	\end{bmatrix} = \frac{\partial {\bf f}}{\partial {\bf q}}
	\label{Jacob}
\end{equation}
equal $\lbrace \pm c(\gamma),0\rbrace$, where the shear wave speed reads
\begin{equation}
	c(\gamma) = \sqrt{1 + 3 \gamma^2 + 1/\tau}\; .
	\label{Speed}
\end{equation}
Since these characteristic wave speeds are always real and distinct, the system \eqref{SystVect} is unconditionally \emph{strictly hyperbolic} \citep{leveque02}.

Note in passing that $\tau \to \infty$ recovers perfect elasticity, and that $\tau\to 0$ recovers strain-rate viscoelasticity. In the latter case, the shear wave speed \eqref{Speed} becomes infinite. This feature is reminiscent of Maxwell--Cattaneo--Vernotte hyperbolic heat conduction which entails infinite wave speeds as the relaxation time tends towards zero, i.e. in the limit of Fourier heat conduction.

\begin{figure}
	\centering
	\includegraphics{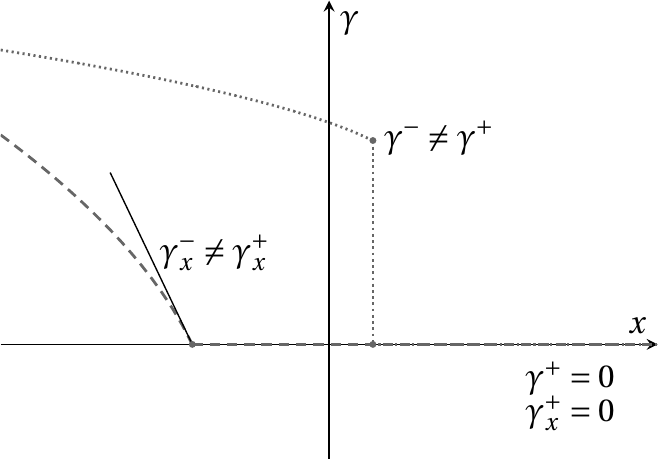}
	
	\caption{Typical wave profile for an acceleration wave (dashed line) and for a shock wave (dotted line) propagating towards increasing $x$ into an undeformed domain. \label{fig:Pic}}
\end{figure}

In the upcoming subsections, we will discuss the properties of two types of singular nonlinear waves, namely \emph{acceleration} waves and \emph{shock} waves. In both cases, we consider right-going waves, i.e., waves that propagate towards increasing $x$, into a region where the material is in a given equilibrium state. In Figure~\ref{fig:Pic}, we represent the typical evolution of the shear strain for acceleration and shock fronts impinging upon an undeformed domain. The analysis of acceleration waves and shock waves will then be used to study travelling wave solutions that keep an invariant profile throughout the motion.

\subsection{Acceleration waves}\label{subsec:ShearAcc}

In a similar fashion to \citet{berjamin22a}, let us analyse the evolution of acceleration waves within the first-order system of balance laws \eqref{SystVect} (i.e., singularities corresponding to jumps in the gradient ${\bf q}_x$ of the primary variables, while the primary variables $\bf q$ remain continuous, see dashed curve in Figure~\ref{fig:Pic}). Thus, we introduce the notation $\llbracket \cdot \rrbracket = (\cdot)^+ - (\cdot)^-$ for the jump of a physical quantity across the surface of discontinuity $x = \varphi(t)$, where $(\cdot)^\pm$ denotes respectively the right- or left-sided values of that quantity. The perturbation is a discontinuity of the gradient ${\bf q}_x$ impinging upon the equilibrium state $\tilde{\bf q}$ with $\tilde r = 0$. The case of a perturbation emerging from an equilibrium state can be analysed in a similar fashion by performing the change of variable $x \to -x$.

\paragraph{Speed.} We consider a right-going acceleration wave travelling at the speed $\Sigma = \dot\varphi(t) > 0$, to be determined. By definition of the jumps and of the surface of discontinuity, the following kinematic relationship holds:
\begin{equation}
	\tfrac{\text d}{\text d t}\llbracket {\bf q}\rrbracket = \big\llbracket \tfrac{\text d}{\text d t} {\bf q}\big\rrbracket = \llbracket {\bf q}_t\rrbracket + \Sigma\, \llbracket {\bf q}_x\rrbracket ,
	\label{Kin}
\end{equation}
where $\frac{\text d}{\text d t} = \partial_t + \Sigma\, \partial_x$ is the directional time derivative along the wavefront. Due to the assumption $\llbracket {\bf q}\rrbracket = {\bf 0}$, the above quantity vanishes, so that the relationship $\llbracket {\bf q}_t\rrbracket = -\Sigma\, \llbracket {\bf q}_x\rrbracket$ is obtained.

Next, evaluation of the jump of \eqref{SystVect} shows that $\Sigma$ is an eigenvalue \eqref{Speed} of the Jacobian matrix ${\bf A}(\tilde{\bf q})$, and that the jump of ${\bf q}_x$ is proportional to the corresponding right eigenvector ${\bf r}$. Therefore, we have $\llbracket {\bf q}_x \rrbracket = \Pi\, {\bf r}$, where
\begin{equation}
	{\bf r} = \begin{bmatrix}
		1\\
		-\Sigma\\
		1/\tau
	\end{bmatrix}, \quad {\bf l} = \frac1{2\Sigma^2}\begin{bmatrix}
		\Sigma^2-1/\tau\\
		-\Sigma\\
		1
	\end{bmatrix} , \quad \Sigma = c(\tilde \gamma) .
	\label{AccVects}
\end{equation}
Here, we have introduced the \emph{wave amplitude} $\Pi = \llbracket \gamma_x \rrbracket$, and the relevant left eigenvector $\bf l$ of ${\bf A}(\tilde{\bf q})$ such that ${\bf l}\cdot {\bf r} = 1$. The sketch in Figure~\ref{fig:Pic} corresponds to the special case $\tilde\gamma = 0$ for which the domain ahead of the wave is not deformed.

\paragraph{Evolution.} In general, the wave amplitude is governed by the Bernoulli differential equation $\dot \Pi = \Omega_2 \Pi^2 - \Omega_1 \Pi$. This result can be obtained by scalar multiplication of \eqref{SystVect} by a left eigenvector, spatial differentiation, and evaluation of the jump across the wavefront, see Section~8.4 of \citet{mueller98}. The coefficients are given by
\begin{equation}
	\begin{aligned}
		\Omega_1 &= -{\bf l}\cdot {\bf R} {\bf r} = \frac1{2\tau^2\Sigma^2} , \\
		\Omega_2 &= -\frac{\partial c}{\partial \bf q} \cdot {\bf r} = -\frac{3\tilde \gamma}{\Sigma} ,
	\end{aligned}
	\label{AccCoeffs}
\end{equation}
where the above formulas were evaluated in the equilibrium state $\tilde{\bf q}$.
Thus, we conclude that steady acceleration waves such that $\dot \Pi = 0$ with nonzero amplitude $\Pi$ might exist, where the wave amplitude is given by the ratio ${\Omega_1}/{\Omega_2} = -\frac16(\tau^2\Sigma\tilde \gamma)^{-1}$. Note in passing that such acceleration waves do not exist in the limit of strain-rate viscoelasticity $\tau \to 0$.

\subsection{Shock waves}\label{subsec:ShearShock}

Now, let us analyse the evolution of shock waves within the first-order system of balance laws \eqref{SystVect} (i.e., strong discontinuities corresponding to jumps in the primary variables $\bf q$, see dotted curve in Figure~\ref{fig:Pic}). Here, we consider a right-going wave that connects the left and right states ${\bf q}^\mp$ located on each side of the shock.

\paragraph{Speed.}

The Rankine--Hugoniot condition for a shock wave with speed $\Sigma = \dot\varphi(t) > 0$ governed by the conservative system \eqref{SystVect} takes the form $\llbracket {\bf f}({\bf q}) \rrbracket = \Sigma\, \llbracket {\bf q} \rrbracket$, see Section 17.12 of \citet{leveque02}. It follows that the shock wave speed satisfies
\begin{equation}
	\Sigma = \sqrt{1 + (\gamma^-)^2 + \gamma^-\gamma^+ + (\gamma^+)^2 + 1/\tau}\; ,
	\label{ShockSpeed}
\end{equation}
where $\gamma^\pm$ are the shear strains to the right and left of the shock. Furthermore, the jumps across the discontinuity satisfy the relationship
\begin{equation}
	\llbracket v \rrbracket = -\Sigma\, \llbracket \gamma \rrbracket = -\tau\Sigma\, \llbracket r \rrbracket .
	\label{RH}
\end{equation}

The existence of shock wave solutions is conditional. Apart from the entropy growth condition \citep{boillat98}, the authors are not aware of a precise shock admissibility criterion for the system \eqref{SystVect}. Nevertheless, one might want to verify the Lax entropy condition \citep{leveque02}, which requires that $c(\gamma^-)>\Sigma>c(\gamma^+)$.
This inequality provides a heuristic criterion for the existence of right-going shock waves. In fact, it reduces to the condition $\gamma^- > \gamma^+$ in the case of non-negative strains.

\paragraph{Evolution.} We adapt the derivation of the results mentioned in \citet{schuler73, chen70} and follow the steps therein. First, we differentiate the first equality of the Rankine--Hugoniot identity \eqref{RH} in time, by means of the product rule. Next, we expand the kinematic relationship \eqref{Kin} for the jumps of $\gamma$ and $v$, and substitutions are performed in the latter based on \eqref{SystAdim}.
This way, we arrive at the general formula
\begin{equation}
	2\sqrt{\Sigma}\; \tfrac{\text d}{\text dt}(\sqrt\Sigma\, \llbracket \gamma \rrbracket) = \Sigma^2 \llbracket\gamma_x \rrbracket - \llbracket (\gamma+\gamma^3+r)_x\rrbracket ,
	\label{ChenGen}
\end{equation}
which corresponds to the identity (2.10) in the study by \citet{chen70}.

The Rankine--Hugoniot relations \eqref{RH} tell us that the jump of the difference $\Delta = \gamma-\tau r$ is uniformly equal to zero. Thus, combining the kinematic relationship \eqref{Kin} for $\Delta$ with \eqref{SystAdim} and \eqref{RH} produces
\begin{equation}
	\llbracket \gamma\rrbracket/\tau + \Sigma\,\llbracket \gamma_x\rrbracket-\tau \Sigma\, \llbracket r_x\rrbracket = 0 ,
	\label{MemShock}
\end{equation}
which allows us to eliminate the memory variable from \eqref{ChenGen}.

Let us now consider a right-going shock wave impinging upon an undeformed material at rest, i.e. the wave speed satisfies \eqref{ShockSpeed} where ${\bf q}^+$ is equal to zero, see Figure~\ref{fig:Pic} for an illustration. In this case, Eqs.~\eqref{ChenGen}-\eqref{MemShock} yield
\begin{equation}
	2\sqrt{\Sigma}\; \tfrac{\text d}{\text dt}\big(\sqrt\Sigma\, \gamma^-\big) = \left(\Sigma^2-c(\gamma^-)^2\right) \left(\gamma_x^- - \gamma_x^*\right) ,
	\label{Chen}
\end{equation}
where
\begin{equation}
	\gamma_x^* = \frac{\gamma^- / (\tau^2\Sigma)}{\Sigma^2-c(\gamma^-)^2} = -\frac{1}{2\tau^2\Sigma\gamma^-}
	\label{GradCrit}
\end{equation}
is a critical strain gradient. We note that the shock wave is steady for $\gamma_x^- = \gamma_x^*$.

\section{Steady progressive waves}\label{sec:Steady}

\subsection{Smooth kinks}\label{subsec:Kinks}

Travelling wave solutions propagate at constant speed without any distortion of the profile. Thus, solutions to \eqref{SystHyp} are sought as functions of the non-dimensional retarded time $\theta = t - x/\mathscr{C}$, where $\mathscr{C} > 0$ represents the wave speed. We emphasise that the neglect of the tensile memory variable $s = s_0(x)\, \text{e}^{-t/\tau}$ after Eq.~\eqref{ShearStress} applies to the present situation as well, given that $s_0 \equiv 0$ is required to express $s$ as a function of the retarded time $\theta$.

The travelling wave Ansatz is injected in \eqref{SystAdim}. Eq.~\eqref{SystAdim}\textsubscript{b} is integrated once with vanishing integration constant after the variable $v$ has been eliminated using \eqref{SystAdim}\textsubscript{a}. The expression of $r$ so-obtained is substituted into \eqref{SystAdim}\textsubscript{c}, leading to the differential equation
\begin{equation}
	\gamma' - \tau \left(\Omega-3\gamma^2\right)\gamma' - \Omega\gamma + \gamma^3 = 0 , 
	\label{KinkDiff}
\end{equation}
for the shear strain, where $\Omega = \mathscr{C}^2-1 \geq 0$. Here, the primes denote differentiation with respect to the retarded time $\theta$.

This differential equation can be simplified further by setting $\gamma = G(\xi) \sqrt{\Omega}$ where $\xi = \Omega\theta$, see e.g. \citet{berjamin23}. This way,
\begin{equation}
	G' = \frac{\Gamma(G)}{1 - \alpha\, \Gamma'(G)} , \qquad \Gamma(G) = G - G^3 ,
	\label{Kink}
\end{equation}
where the prime denotes differentiation with respect to the argument (here, the explicit dependency of $G$ on $\xi$ has been omitted), and $\alpha = \Omega \tau \geq 0$ is a parameter. This differential equation is analogous to Eq.~(5.6) of \citet{pucci15}.

From Eq. \eqref{Kink}, one observes that travelling wave solutions to the hyperbolic system \eqref{SystHyp} should connect the equilibrium strains $G = 0$ and $G = \pm 1$ by following a smooth transition that depends on the parameter $\alpha$. Solutions are obtained by rewriting \eqref{Kink} in separable form, and by performing a partial fraction decomposition. Upon integration, we find
\begin{equation}
	\xi = \ln \left(\frac{3^{\alpha} \sqrt{3}}{8^{\alpha}} \frac{G^{1-\alpha}}{(1-G^2)^{1/2+\alpha}}\right)
	\label{KinkSol}
\end{equation}
in implicit form, where we have enforced $G(0)=\frac12$ without loss of generality. The waveform can be written in explicit form when $\alpha = 0, \frac14, 1$, in which cases $G(\xi)$ equals
\begin{equation}
	\frac{\text{e}^{\xi}}{\sqrt{3+\text{e}^{2\xi}}} ,\;
	\frac{4\, \text{e}^{4\xi/3}}{3 + \sqrt{9 + 16\, \text{e}^{8\xi/3}}}, \;
	\sqrt{1 - \tfrac34 \text{e}^{-2\xi /3}} ,
	\label{KinkSolsPart}
\end{equation}
respectively.

\begin{figure*}
	\centering
	\includegraphics{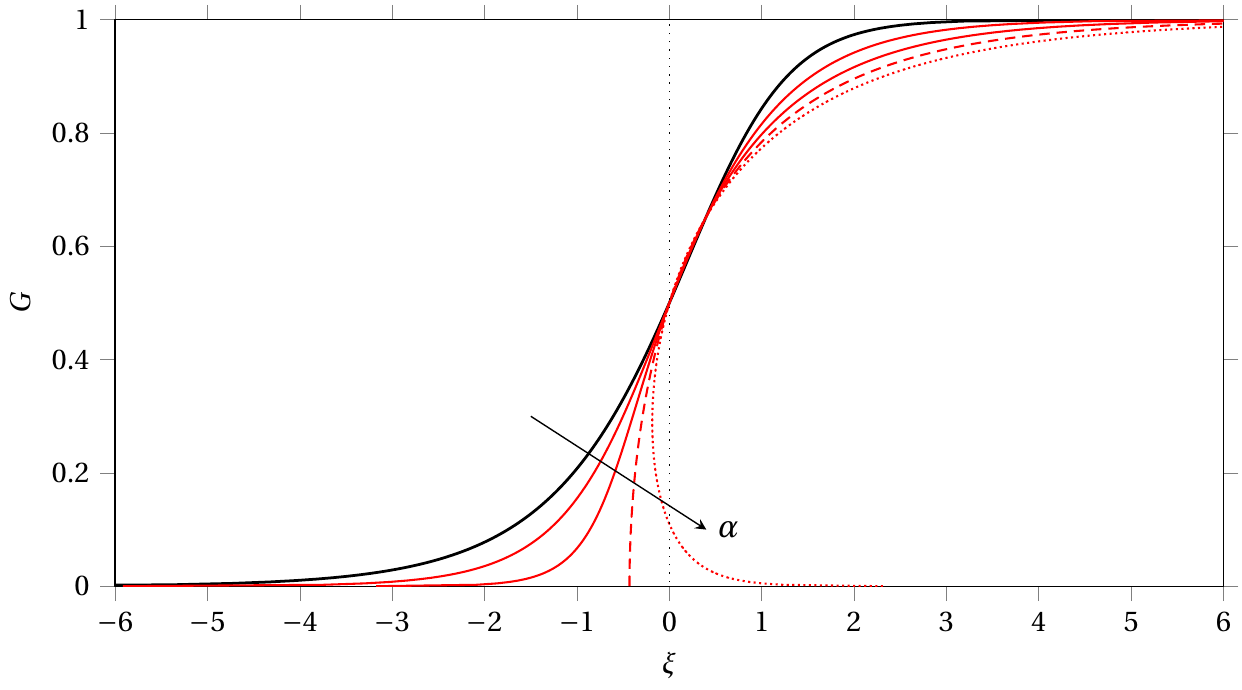}
	
	\caption{ Smooth kinks \eqref{KinkSol}. Wave profile in terms of the rescaled retarded time for $\alpha = 0, \frac13, \frac23, 1, \frac43$ (increasing values marked by the arrow). Multivalued solutions are obtained for $\alpha > 1$. \label{fig:KinkVar} }
\end{figure*}

Figure~\ref{fig:KinkVar} displays the kink waveforms obtained for several values of the parameter $\alpha$. We note that the smoothness of the solution is impacted when $\alpha$ approaches one: the slope $G'$ of the curve becomes infinite locally, while $G$ remains continuous. The smooth kink solution even becomes multi-valued for $\alpha > 1$, thus revealing the presence of a shock. This observation is consistent with anterior works \citep{pucci15}.

\begin{figure}
	\centering
	\includegraphics{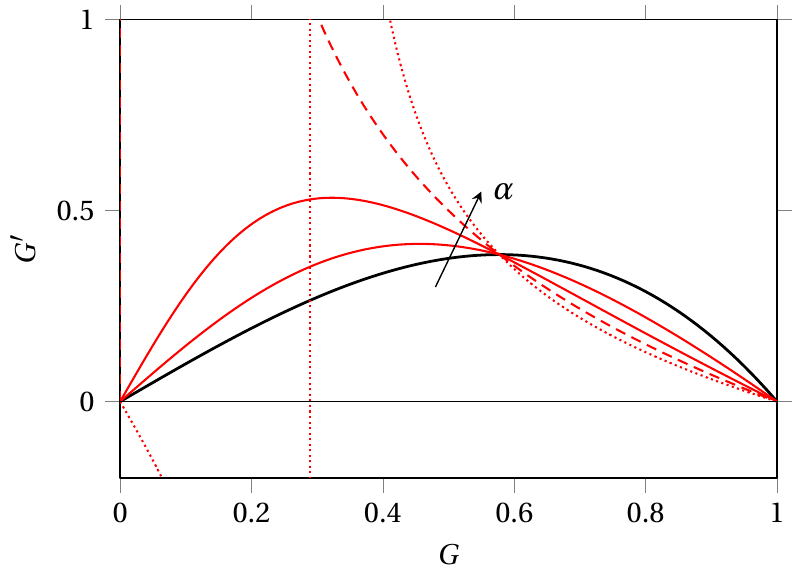}
	
	\caption{ Trajectories of the kink solution \eqref{KinkSol} for $\alpha = 0, \frac13, \frac23, 1, \frac43$. A singularity is obtained for $\alpha > 1$. \label{fig:KinkPh} }
\end{figure}

Figure~\ref{fig:KinkPh} provides an alternative representation of this phenomenon. Here, we display the evolution \eqref{Kink} of $G'$ in terms of $G$ for the same values of $\alpha$ as in Figure~\ref{fig:KinkVar}. For $\alpha = \frac43$, a vertical asymptote located at $G \approx 0.289$ divides the plane into two regions. This way, no smooth trajectory along which $G'$ remains finite can connect $G=0$ and $G=1$.

\subsection{Steady acceleration waves}\label{subsec:SteadyAcc}

We go back to the study of acceleration waves, see Section~\ref{subsec:ShearAcc}. If the singularity travels at the same speed as the kink, then $\mathscr{C}$ must be equal to the characteristic speed  $c(\tilde \gamma)$ in the equilibrium state $\tilde{\bf q}$. The wave is steady if its amplitude $\Pi = \llbracket \gamma_x \rrbracket$ is equal to the ratio $\Omega_1/\Omega_2 = -\frac16(\tau^2\Sigma\tilde \gamma)^{-1}$ deduced from the coefficients in Eq.~\eqref{AccCoeffs}.

In terms of the rescaled variables used for the kinks, the strain gradient satisfies $-\mathscr{C}\gamma_x \Omega^{-3/2} = G'$. Thus, by computing the jumps on each side of this equality, the expression of $\llbracket G'\rrbracket$ is deduced from the condition $\Pi = \Omega_1/\Omega_2$. Since the wave speed satisfies $\mathscr{C} = c(\tilde G \sqrt{\Omega})$, we deduce from \eqref{Speed} that the acceleration wave is steady if
\begin{equation}
	\llbracket G'\rrbracket = \frac{1}{6\alpha^2 \tilde G}, \qquad
	\tilde G = \sqrt\frac{\alpha-1}{3\alpha} .
	\label{AccSteady}
\end{equation}
The above equilibrium strain is in the range $0 \leq \tilde G< 1/{\sqrt 3}$ for all $\alpha \geq 1$. At the critical value $\alpha = 1$, the equilibrium strain $\tilde G$ vanishes, and the jump $\llbracket G'\rrbracket$ becomes infinite. Therefore, a vertical asymptote is obtained, as illustrated in Figure~\ref{fig:KinkVar} (dashed line).

\subsection{Steady shocks}\label{subsec:SteadyShocks}

We refer to Section~\ref{subsec:ShearShock} for the study of shock waves entering an undeformed domain at rest. Since the singularity travels at the same speed as the kink, the wave velocity $\mathscr{C}$ must be equal to the shock speed $\Sigma$ of Eq.~\eqref{ShockSpeed}. The wave is steady if the strain gradient $\gamma_x^-$ before the shock is equal to the critical value $\gamma_x^*$ in Eq.~\eqref{GradCrit}.

In terms of the rescaled variables, the definition of the coordinate $\xi$ in Section~\ref{subsec:Kinks} leads to an inversion of left and right sides, that is $\gamma^- = G^+\sqrt{\Omega}$, etc. The critical strain gradient \eqref{GradCrit} satisfies $-\mathscr{C}\gamma_x^* \Omega^{-3/2} = G^{\prime +}$, while the shock wave speed is deduced from $\mathscr{C} = \Sigma$ with $\gamma^- = G^+\sqrt{\Omega}$ and $\gamma^+=0$. Therefore, the wave is steady if
\begin{equation}
	G^{\prime +} = \frac{1}{2\alpha^2G^+} , \qquad
	G^+ = \sqrt\frac{\alpha-1}{\alpha} .
	\label{ShockSteady}
\end{equation}
The above value of the strain jump is in the range $0 < G^+< 1$ for all $\alpha > 1$.

The strain gradient $G^{\prime +}$ ensuring shock stability is compatible with the differential equation \eqref{Kink} evaluated at $G^+$. Therefore a stable shock wave can be connected to the smooth kink, under the conditions \eqref{ShockSteady}. It suffices to truncate the multivalued smooth waveform \eqref{KinkSol} at the coordinate $\xi$ where the solution reaches the critical value $G^+$.

\begin{figure}
	\centering
	\includegraphics{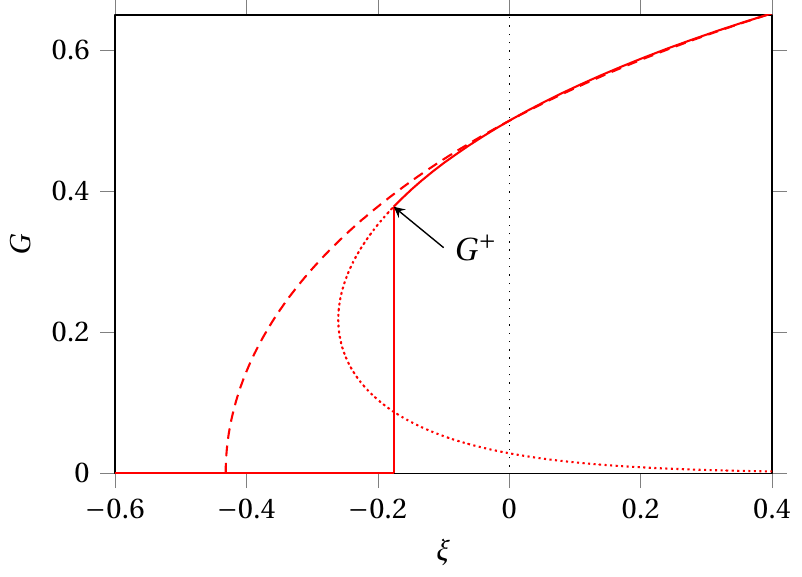}
	
	\caption{Singular kinks \eqref{KinkSol} governed by \eqref{AccSteady} and \eqref{ShockSteady}. Evolution of the normalised strain in terms of the rescaled retarded time for $\alpha = \frac76$ where a steady shock develops (solid line). The dashed line marks the critical case $\alpha = 1$. \label{fig:KinkShock} }
\end{figure}

This procedure is illustrated in Figure~\ref{fig:KinkShock} where we have set $\alpha = \frac76$. Thus, according to Eq.~\eqref{ShockSteady}\textsubscript{b}, the amplitude of the discontinuity is given by $G^+ \approx 0.378$. The velocity of the shock is deduced from the relationship $\alpha = \Omega \tau$ where $\Omega = \mathscr{C}^2-1$. In other words, we find $\mathscr{C} = \sqrt{1+\alpha/\tau}$ where $1/\alpha = 1-(G^+)^2$. The relationship between velocity and amplitude is further explored in the next subsection, including its implications in terms of the initial physical variables.

\subsection{Amplitude-velocity relationship}

We note that the maximum amplitude $\sup_\xi(G)=1$ of a steady progressive wave travelling at speed $\mathscr{C}$ does not depend on $\alpha$. However, the amplitude $G^+$ of the discontinuity depends on $\alpha$ according to \eqref{ShockSteady}\textsubscript{b}. This property is illustrated hereinafter.

For this purpose, we express the previous results in terms of the dimensionless variables and parameters of \eqref{SystHypAdim}-\eqref{SystAdim}. For a unit kink covering the range $0\leq G\leq 1$, the corresponding shear strains $\gamma$ are bounded by the maximum value $\gamma_{\sup} = \sqrt{\Omega}$ where $\Omega = \mathscr{C}^2 - 1$. Therefore, the maximum strain amplitude does not depend on $\tau$, but the maximum strain is directly connected to the wave speed.

The condition $\alpha > 1$ ensuring the formation of a steady shock can be expressed in terms of the maximum strain amplitude as $\gamma_{\sup} > \sqrt{1/\tau}$. Therefore, singular travelling waves may arise only beyond a critical dimensionless strain, whose magnitude decreases with increasing values of the dimensionless relaxation time $\tau$, see the notations of Eq.~\eqref{SystAdim}. For such a discontinuous wave, the amplitude of the discontinuity $\gamma^-$ is linked to the wave speed $\mathscr{C}$ according to Eq.~\eqref{ShockSpeed}, which involves the parameter $\tau$ explicitly.

\begin{figure}
	\centering
	\includegraphics{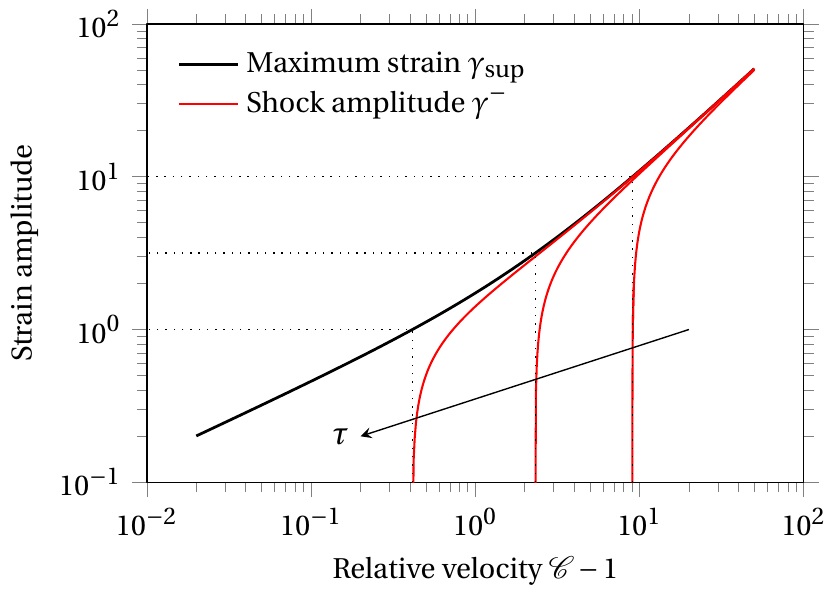}
	
	\caption{Amplitude-velocity relationship for travelling waves \eqref{KinkSol}-\eqref{ShockSteady}. Evolution of the maximum dimensionless strain $\sup (\gamma)$ and of the shock amplitude $\gamma^-$ with respect to the velocity for several values of $\tau = 0.01, 0.1, 1$ (arrow marking increasing values). \label{fig:Speed}}
\end{figure}

Figure~\ref{fig:Speed} represents the evolution of the maximum strain $\gamma_{\sup}$ and of the shock amplitude $\gamma^-$ in terms of the relative wave speed $\mathscr{C}-1$ for several values of $\tau$. The condition $\gamma_{\sup} > \sqrt{1/\tau}$ leading to the existence of steady shocks is marked by dotted lines. While the amplitude-velocity relationship for the maximum strain is unaffected by variations of the parameter $\tau$, the shock wave amplitude is very sensitive to such variations.

In terms of the initial variables used in Eq.~\eqref{SystHyp}, the physical shear strain satisfies
\begin{equation}
	\gamma = G(\xi) \sqrt{\frac{\Omega}{b}} , \qquad \xi = \frac{\alpha}{\tau} \bigg(t-\frac{x \sqrt{\rho/\mu}}{\sqrt{1+\Omega}}\bigg) ,
\end{equation}
where we have used the definition of $\Omega = \alpha\eta/(\tau\mu)$ and the rescaling defined in \eqref{SystHypAdim}. Therefore, for a given set of physical material parameters $\rho$, $\mu$, $b$, $\eta$, $\tau$, the waveform is completely determined by the provision of the parameter $\alpha$. In fact, given that $0\leq G\leq 1$, we note that the physical strain amplitude $\gamma_{\sup} = \sqrt{\Omega/b}$ increases with increasing values of $\alpha$. The physical wave speed $\sqrt{(1+\Omega) \mu/\rho}$ increases with $\alpha$ as well. Moreover, this quantity can be connected to the strain amplitude, and we arrive at the expression $1+b \gamma_{\sup}^2$ for the squared wave speed multiplied by $\rho/\mu$. This equation provides a means to determine the value of the parameter of nonlinearity $b$, but it does not involve the viscoelastic parameters.

According to the discussions in the previous paragraphs, discontinuous travelling waves might be used to determine the viscoelastic parameters in a similar fashion. In fact, the squared physical shock velocity multiplied by $\rho/\mu$ equals $1 + b(\gamma^-)^2 + \eta/(\mu\tau)$, which involves the parameters $\eta$, $\tau$ explicitly.

Above observations highlight the influence of the  viscoelastic parameters for the existence of steady shocks when one single relaxation mechanism is considered. While the material parameters cannot be adjusted easily in practice, we emphasise that real materials are better described by a sequence of relaxation mechanisms covering a wide range of relaxation frequencies, see \citet{carcione15}. Thus, for any imposed strain amplitude, one of the numerous relaxation times might be large enough for the corresponding relaxation mechanism to contribute to the development of singularities in the wave profile on its own. Nevertheless, it remains to investigate the interplay between relaxation mechanisms with distinct characteristic times before a more conclusive statement can be made.

\section{Multiple relaxation mechanisms}\label{sec:Multiple}

\subsection{Governing equations}

If two relaxation mechanisms are considered, then the constitutive law \eqref{Constitutive}-\eqref{ConstitVisc} in Section~\ref{sec:GovEq} is modified. In this case, a sum of two decaying exponentials is inserted in Eq.~\eqref{Memory}, with the parameters $\eta_\ell$, $\tau_\ell$ for $\ell = 1,2$ \citep{haupt02}. This procedure results in an additive decomposition of the viscous stress,
\begin{equation}
	\begin{aligned}
		&\bm{T}^\text{v} = \bm{T}^\text{v}_1 + \bm{T}^\text{v}_2 , \\
		&\bm{T}^{\text{v}}_\ell + \tau_\ell\, \mathscr{D}(\bm{T}^{\text{v}}_\ell) =  2\eta_\ell \bm{D} , \qquad (\ell = 1,2) .
	\end{aligned}
	\label{Multiple}
\end{equation}
This decomposition involves two partial viscous stress tensors $\bm{T}^{\text{v}}_\ell$ governed by similar evolution equations as in the previous case \eqref{ConstitVisc}, which is recovered for $\bm{T}^{\text{v}}_2 \equiv \bm{0}$. Generalisation to an arbitrary number of relaxation mechanisms is straightforward.

Let us linearise the equations of motion \eqref{EqMot} about an undeformed motionless state. Doing so, we note that the linearised elastic stress \eqref{LinElast} is unchanged when multiple relaxation mechanisms are considered, and that the linearised viscous stress \eqref{Multiple}\textsubscript{a} is still decomposed additively. The partial viscous stress tensors $\bm{T}^{\text{v}}_\ell$ in \eqref{Multiple}\textsubscript{b} are now governed by linear evolution equations which are similar to Eq.~\eqref{LinVisc}.

Next, we form a specific linear combination of $\bm T$ and its time derivatives. This way, we arrive at the following differential equation for the deviatoric stress,
\begin{equation}
	{\bm T}_\text{d} + \tau \dot{\bm T}_\text{d} + \nu^2 \ddot{\bm T}_\text{d} = 2\mu \left(\bm{\varepsilon}_\text{d} + \tau_\varepsilon \dot{\bm \varepsilon}_\text{d} + \nu_\varepsilon^2 \ddot{\bm \varepsilon}_\text{d} \right),
	\label{SLSMult}
\end{equation}
with the coefficients $\tau = \tau_1 + \tau_2$ and $\nu^2 = \tau_1\tau_2 < \tau^2/4$, as well as
\begin{equation}
	\begin{aligned}
		&\tau_\varepsilon = \tau + \eta/\mu ,  \qquad
		\nu_\varepsilon^2/\nu^2 = 1  + \eta \langle \tau^{-1} \rangle /\mu .
	\end{aligned}
	\label{SLSMultCoeff1}
\end{equation}
The total viscosity reads $\eta = \eta_1+\eta_2$, and the average relaxation frequency is defined as
\begin{equation}
	\langle \tau^{-1} \rangle = \zeta_1/\tau_1 + \zeta_2/\tau_2, \qquad \zeta_\ell = \eta_\ell/\eta .
	\label{SLSMultCoeff2}
\end{equation}
Dispersion analysis provides the relationship
\begin{equation}
	\rho \frac{\omega^2}{\kappa^2} = \frac{\mu}{2} \left(\frac{1+\text{i}\omega \tau_{\varepsilon 1}}{1+\text{i}\omega \tau_1} + \frac{1+\text{i}\omega \tau_{\varepsilon 2}}{1+\text{i}\omega \tau_2}\right),
	\label{DispersionSLSMult}
\end{equation}
where $\tau_{\varepsilon \ell} = \tau_\ell + 2\eta_\ell/\mu$, which is a generalisation of the SLS dispersion relationship \eqref{DispersionSLS}.

Similarly to Section~\ref{sec:Shear}, we restrict the analysis to simple shear motions \eqref{SimpleShear}, where the memory variable $r = r_1 + r_2$ defined in Eq.~\eqref{MemVars} is decomposed additively. Here, the reference coordinates $t_\text{r}$, $x_\text{r}$ used to rescale the time coordinates involve the total viscosity $\eta$. Using these definitions, we introduce the same dimensionless strains and velocities as in Eq.~\eqref{SystHypAdim}, as well as a similar scaling for the relaxation times $\tau_\ell$. Thus, for $\phi = \pm 1$, we arrive at
\begin{equation}
	\left\lbrace
	\begin{aligned}
		\gamma_{t} &= v_{x} ,\\
		v_{t} &= (\gamma + \gamma^3 + r_1 + r_2)_{x} , \\
		\tau_\ell (r_\ell)_t &= \zeta_\ell v_{x} - r_\ell , \qquad (\ell = 1,2) .
	\end{aligned}
	\right.
	\label{SystAdimMult}
\end{equation}
Again, overbars have been omitted. This system can be written in the form of a hyperbolic system of balance laws \eqref{SystVect} in terms of the variables ${\bf q} = [\gamma, v, r_1, r_2]^\top$.

\subsection{Acceleration waves}

If we reproduce the steps of Section~\ref{subsec:ShearAcc}, we find that the speed of a right-going acceleration wave is
\begin{equation}
	\Sigma = c(\tilde\gamma) = \sqrt{1+3\gamma^2+\langle \tau^{-1} \rangle}\; .
	\label{SpeedMult}
\end{equation}
The coefficients governing the evolution of the slope in \eqref{AccCoeffs} become
\begin{equation}
	\Omega_1 = \frac{\langle \tau^{-2} \rangle}{2\ \Sigma^2} , \quad
	\Omega_2 = -\frac{3\tilde \gamma}{\Sigma} ,
	\label{CoeffMult}
\end{equation}
where $\langle \tau^{-2} \rangle = {\zeta_1}/{\tau_1}^2+{\zeta_2}/{\tau_2}^2$.
Therefore, steady acceleration waves can be obtained in the present case as well.

\subsection{Shock waves}

If we reproduce the steps of Section~\ref{subsec:ShearShock}, we find that the speed of a right-going shock wave is
\begin{equation}
	\Sigma = \sqrt{1 + (\gamma^-)^2 + \gamma^-\gamma^+ + (\gamma^+)^2 + \langle \tau^{-1} \rangle}\; ,
	\label{ShockSpeedMult}
\end{equation}
according to the Rankine--Hugoniot shock conditions. In particular, the relationships
\begin{equation}
	\llbracket v\rrbracket = -\Sigma\, \llbracket \gamma \rrbracket = -\tau_\ell \Sigma\, \llbracket r_\ell\rrbracket / \zeta_\ell
\end{equation} with $\ell = 1,2$ hold across the discontinuity.

The evolution of the shock amplitude is still governed by Eq.~\eqref{ChenGen}. Here, we note that the quantities $\Delta_\ell = \gamma - \tau_\ell r_\ell/\zeta_\ell$ are continuous across the wavefront. Thus, by following similar steps, we recover \eqref{Chen} with the critical strain gradient
\begin{equation}
	\gamma_x^* = \frac{\langle \tau^{-2} \rangle}{\Sigma^2 - c(\gamma^-)^2}\frac{\gamma^-}{\Sigma} = -\frac{\langle \tau^{-2} \rangle}{2 \Sigma \gamma^-} ,
	\label{GradCritMult}
\end{equation}
where $c(\gamma^-)$, $\Sigma$ are deduced from Eqs.~\eqref{SpeedMult}-\eqref{ShockSpeedMult}, respectively. This equation generalises Eq.~\eqref{GradCrit}.

\subsection{Smooth kinks}

We proceed in a similar fashion to Section~\ref{subsec:Kinks}. Similarly to the derivation of \eqref{KinkDiff}, we combine Eqs.~\eqref{SystAdimMult}\textsubscript{a}-\eqref{SystAdimMult}\textsubscript{c} to arrive at a first-order differential equation for $\gamma$ that still depends on $r_2$. Substitution of $r_2$ in Eq.~\eqref{SystAdimMult}\textsubscript{d} then yields the differential equation
\begin{equation}
	a_0(G) G'' = \Gamma(G) + a_1(G) G' + a_3(G) (G')^2 , 
	\label{KinkMult}
\end{equation}
where $\Gamma(G) = G-G^{3}$ and
\begin{equation}
	\begin{aligned}
		a_0(G) &= \delta^2 \left(\langle \tau^{-1}\rangle/\Omega - \Gamma'(G) \right) ,\\
		a_1(G) &= \alpha\, \Gamma'(G) - 1 , \qquad
		a_3(G) = \delta^2 \Gamma''(G) .
	\end{aligned}
	\label{KinkMultCoeff}
\end{equation}
Here, $\alpha = \Omega \tau$ and $\delta = \Omega \nu$, see also the coefficients defined in Eqs.~\eqref{SLSMultCoeff1}-\eqref{SLSMultCoeff2}. We note that Eq.~\eqref{Kink} is recovered for $\delta = 0$, and that smooth solutions to \eqref{KinkMult} should connect the equilibrium strains $G = 0$ and $G = \pm 1$.

Fully analytical expressions seem difficult to obtain when multiple relaxation mechanisms are considered. For $0< \delta < \alpha/2$, travelling wave solutions can be computed numerically by integration of \eqref{KinkMult} based on a suitable numerical method. By choosing the initial data such that $G(0) = \frac12$ and $G'(0)$ is deduced from Eq.~\eqref{Kink}, we did not manage to find a range of parameter values that leads to smooth solutions, numerically. From these simulations, we conclude that the emergence of singular travelling wave solutions is very likely. Unfortunately, the full analytical derivations carried out in Section~\ref{sec:Shear} can hardly be reproduced in a straightforward manner in the present case. We can expect that the complexity of the governing equations increases when an arbitrary number $N>2$ of relaxation mechanisms is considered.

\section{The Jaumann model}\label{sec:Jaumann}

\subsection{Governing equations}

We now consider the system \eqref{SystHyp} with the Jaumann objective rate $\phi = 0$. In non-dimensional form \eqref{SystHypAdim}, we thus have
\begin{equation}
	\left\lbrace
	\begin{aligned}
		\gamma_{t} &= v_{x} ,\\
		v_{t} &= (\gamma + \gamma^3 + r)_{x} , \\
		\tau r_{t} &= (1+\tau_b s) v_{x} - r ,\\
		\tau s_{t} &= -\tau_b r v_{x} - s ,
	\end{aligned}
	\right.
	\label{SystJaumann}
\end{equation}
where we have introduced $\bar \tau_b = \bar \tau/\sqrt{b}$. Again, overbars have been omitted in \eqref{SystJaumann} for the sake of simplicity. This way, the system \eqref{SystAdim} studied in the previous sections is recovered for $\tau_b = 0$.

The first-order system of partial differential equations \eqref{SystJaumann} can be written in quasi-linear form
\begin{equation}
	{\bf q}_t + {\bf A}({\bf q}){\bf q}_x = {\bf R} {\bf q} ,
	\label{SystJVect}
\end{equation}
where ${\bf q} = [\gamma,v,r,s]^\top$,
\begin{equation}
	{\bf A}({\bf q}) = -\begin{bmatrix}
		0 & 1 & 0 & 0\\
		1 + 3\gamma^2 & 0 & 1 & 0\\
		0 & (1+\tau_b s)/\tau & 0 & 0\\
		0 & -\tau_b r/\tau & 0 & 0
	\end{bmatrix} ,
	\label{SystJVar}
\end{equation}
and
\begin{equation}
	{\bf R} = - \begin{bmatrix}
		0 & 0 & 0 & 0\\
		0 & 0 & 0 & 0\\
		0 & 0 & 1/\tau & 0\\
		0 & 0 & 0 & 1/\tau
	\end{bmatrix} .
\end{equation}
Here, the eigenvalues of the diagonalisable matrix ${\bf A}({\bf q})$ equal $\lbrace \pm c(\gamma),0,0\rbrace$, where the shear wave speed satisfies
\begin{equation}
	c(\gamma,s) = \sqrt{1 + 3 \gamma^2 + (1+\tau_b s)/\tau} \; .
	\label{SpeedJ}
\end{equation}
Since the wave speed could possibly become complex at large $|s|$ unless $\tau_b = 0$, the system \eqref{SystJVect} is only conditionally \emph{hyperbolic}.

In theory, waves do no longer propagate if hyperbolicity is lost, thus restricting the validity of the present results. In practice, the memory variable $s$ in Eqs.~\eqref{SystJaumann}-\eqref{SpeedJ} might not be able to take arbitrarily large values, even though this outcome does not seem totally impossible at first sight. Furthermore, a loss of hyperbolicity might only happen at very large strains where plastic behaviour or other physics must be accounted for.

\subsection{Acceleration waves}

If we reproduce the steps of Section~\ref{subsec:ShearAcc}, we find that the speed of a right-going acceleration wave is $\Sigma = c(\tilde\gamma, 0)$, where $\tilde r$, $\tilde s$ vanish in the equilibrium state $\tilde{\bf q}$. Thus, the speed of such an acceleration wave is the same as in Eq.~\eqref{AccVects}, see expression in Eq.~\eqref{SpeedJ} with $s = 0$. At equilibrium, the coefficients $\Omega_1$, $\Omega_2$ governing the evolution of the slope in Eq.~\eqref{AccCoeffs} are unchanged too (note that these expressions were evaluated in the equilibrium state $\tilde{\bf q}$).
Therefore, steady acceleration waves can be obtained in the present case as well.

\subsection{Shock waves}

Unfortunately, the steps in Section~\ref{subsec:ShearShock} can hardly be reproduced for the system \eqref{SystJaumann} due to the presence of non-conservative products $s v_x$ and $r v_x$. Hence, the study of shear shock waves is less straightforward in this case, starting with the estimation of their speed \citep{lefloch89,camacho08}. To address this case, one might seek a change of variables that yields a conservative form of the equations of motion. Otherwise, analytical approximations may prove useful, see \citet{fu90}.

\subsection{Smooth kinks}

We proceed in a similar fashion to Section~\ref{subsec:Kinks}. Similarly to the derivation of \eqref{KinkDiff}, we combine Eqs.~\eqref{SystJaumann}\textsubscript{a}-\eqref{SystJaumann}\textsubscript{c} to arrive at a first-order differential equation for $\gamma$ that still depends on $s$. Substitution of $s$ in Eq.~\eqref{SystJaumann}\textsubscript{d} then yields the differential equation
\begin{equation}
	\alpha\, \Gamma(G) \frac{G''}{G'} = \Gamma(G) + a_3(G) G' + a_4(G) (G')^2, 
	\label{KinkJ}
\end{equation}
where $\Gamma(G) = G-G^{3}$ and
\begin{equation}
	\begin{aligned}
		a_3(G) &= 2\alpha\, \Gamma'(G) - 1 ,\\
		a_4(G) &= \alpha^2 \left(\delta^{2} \Gamma(G) + \Gamma''(G)\right) .
	\end{aligned}
	\label{KinkJCoeffs}
\end{equation}
Here, we have introduced the coefficient $\delta = \Omega^{3/2}\tau_b/\alpha$ (other notations are unchanged). This differential equation is of the same type than the one studied by \citet{depascalis19}.

Under the condition that $G''/G' \to 0$ near equilibrium, we find that $\Gamma(G) = 0$ at equilibrium. Thus, most travelling wave solutions to the hyperbolic system \eqref{SystHyp} should connect the equilibrium strains $G = 0$ and $G = \pm 1$ by following a smooth transition. For $\alpha> 0$ and $\delta = 0$, the second-order differential equation \eqref{KinkJ} reduces to the case of Section~\ref{subsec:Kinks} (upon differentiation of \eqref{Kink} with respect to $\xi$), for which analytical solutions are obtained in implicit form.

For $\delta > 0$, travelling wave solutions can be computed numerically by integration of \eqref{KinkJ} based on Matlab's \texttt{ode15s} solver, or another similar method. Since the differential equation \eqref{KinkJ} is of second order, resolution of the initial-value problem requires that the values of $G(0)$ and $G'(0)$ are provided. However, not every choice of this kind will necessarily make $G''/G'$ decay at large rescaled times. Here, the starting values are chosen such that $G(0) = \frac12$, and $G'(0)$ minimises the quantity $|G(-10)| + |1-G(10)|$ found numerically. For this optimisation, we use Matlab's \texttt{fminsearch} function, where the initial guess for $G'(0)$ is deduced from Eq.~\eqref{Kink} with $G=\frac12$.

\begin{figure}
	\centering
	\includegraphics{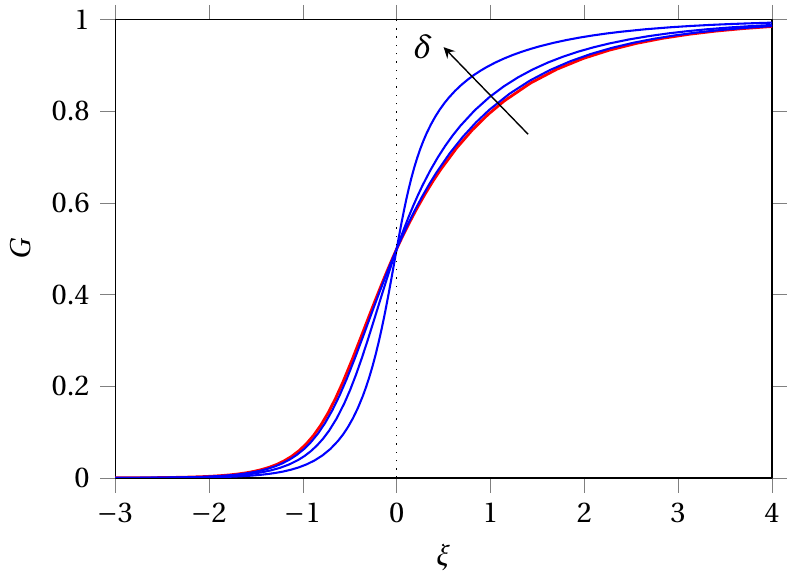}
	
	\caption{Jaumann model. Smooth kinks \eqref{KinkJ} obtained numerically for $\alpha = \frac23$ and several values of $\delta\geq 0$ (the arrow marks increasing values). \label{fig:Jaumann}}
\end{figure}

Figure~\ref{fig:Jaumann} represents the solutions so-obtained for $\alpha = \frac23$ and for several values of $\delta = 0,1,2,3$ corresponding to the initial strain gradients $G'(0) \approx 0.45, 0.48, 0.59, 1.15$. From these simulations, we observe that increasing values of the parameter $\delta$ have a wavefront stiffening effect, where $\delta=0$ corresponds to the curve displayed in Figure~\ref{fig:KinkVar}. Observations made in the upper- and lower-convected cases $\phi=\pm 1$ suggest that steady shock wave solutions might exist in the Jaumann case $\phi=0$ as well.

\section{Concluding remarks}\label{sec:Conclusion}

We investigated the propagation of travelling shear waves within a class of rate-type viscoelasticity models of soft solid. The models based on the lower- and upper-convected Cauchy stress rates provide useful simplifications. Within this framework, we studied the propagation of steady progressive waves, which can include discontinuous parts in some cases. Closed form analytical expressions are obtained, both in the case of smooth wave profiles and of singular ones.

However, the study of shear shock waves is less straightforward in the case of multiple relaxation mechanisms, as well as for stress rate models involving the Jaumann derivative. This is due to the emergence of higher-order differential equations, or to the presence of non-conservative products in the first-order system \eqref{SystHyp} that would necessitate a suitable change of variables.

Let us reflect on possible constitutive modelling choices to reproduce the nonlinear viscoelastic behaviour of soft solids, including creep and stress relaxation. In the spirit of Fung's quasi-linear viscoelasticity theory (QLV), a first modelling approach requires very little information \citep{berjamin22a}. In fact, the viscoelastic behaviour at large strains is directly deduced from the nonlinear elastic response at long times, and from time-dependent relaxation measurements at small strains.

As shown in the Appendix~\ref{sec:Appendix}, the present modelling approach based on the works by \citet{haupt02} and \citet{saccomandi21} is the most natural formulation of its kind, but it is slightly more complex than QLV. One advantage is that the constitutive law involves only objective quantities, and that Newtonian viscosity theories are included in the model as a special case. However, one drawback is the need to select one objective stress rate \eqref{ObjectiveDer} among several options. The most suitable parameter $\phi$ could be determined from the measurement of compression stresses in simple shearing motions, see Eq.~\eqref{SystHyp}. Alternatively, one might consider uniaxial tensile motions, see Appendix~\ref{sec:Appendix}. The fact that upper- and lower-convected rates simplify calculations is not a valid argument if laboratory experiments prove otherwise.

%\printcredits
%
%\section*{Declaration of competing interest}
%
%The authors declare that they have no known competing financial interests or personal relationships that could have appeared to influence the work reported in this paper.

\section*{Acknowledgements}

HB has received funding from the European Union's Horizon 2020 research and innovation programme under grant agreement TBI-WAVES{\,--\,}H2020-MSCA-IF-2020 project No 101023950. GS is partially supported by the PRIN research project `The mathematics and mechanics of nonlinear wave propagation in solids' (grant no. 2022P5R22A), and by GNFM of Istituto Nazionale di Alta Matematica (INDAM).

\appendix

\section{Comment on \citet{farina22}}\label{sec:Appendix}

The tensile creep problem considered by \citet{farina22} for Mooney--Rivlin materials of rate type highlights physical inconsistencies when $\phi = \pm 1$ (`third model' therein). More specifically, additional restrictions on the model parameters are proposed to ensure dissipative behaviour. This observation seems contradictory to the thermodynamic consistency property established by \citet{haupt02}. To overcome this paradox, we reconsider the tensile creep problem hereinafter. Contrary to the discussions in the introduction of \citet{farina22}, we emphasise that thermodynamic consistency is crucial out of equilibrium, even for the study of `isothermal processes'.

Let us assume that the deformation gradient tensor is diagonal and of unit determinant. Its uniaxial component $F_{11} = \lambda$ is the tensile stretch. Consequently, the stress tensor $\bm T$ is diagonal, and the lateral stresses are equal, $T_{22} = T_{33}$. According to the above constitutive equations \eqref{Constitutive}-\eqref{ConstitElast} with $\beta=0$ (Mooney--Rivlin material), we obtain the following components of the elastic stress:
\begin{equation}
	\begin{aligned}
		T_{11}^\text{e} &= 2 C_1 \lambda^2 - 2 C_2 \lambda^{-2} ,\\
		T_{22}^\text{e} &= 2C_1 \lambda^{-1} - 2 C_2 \lambda .
	\end{aligned}
	\label{StressComp}
\end{equation}
Setting the total uniaxial stress $T_{11}$ constant and making the lateral stresses vanish, we arrive at $T_{11} = \sigma^\text{e} + \sigma^\text{v}$ with $\sigma^{\bullet} = T_{11}^{\bullet} - T_{22}^{\bullet}$, by substitution of the pressure. In particular, the elastic stress contribution satisfies
\begin{equation}
	\begin{aligned}
		\sigma^\text{e} &= 2 \left(C_1 \lambda + C_2\right) \left(\lambda - \lambda^{-2}\right) ,\\
		\sigma^{\text{e}\prime} &= 2 C_1 \left(2\lambda + \lambda^{-2}\right) + 2 C_2 \left(1 + 2 \lambda^{-3}\right) ,
	\end{aligned}
	\label{StressElast}
\end{equation}
where the prime $^\prime$ denotes differentiation with respect to $\lambda$. The evolution of the viscous stress components \eqref{ConstitVisc}-\eqref{ObjectiveDer} entails
\begin{equation}
	\begin{aligned}
		\left(1 + 2\phi\tau \dot\Lambda\right) T_{11}^\text{v} + \tau \dot T_{11}^\text{v} &= 2\eta \dot\Lambda ,
		\\
		\left(1-\phi\tau\dot\Lambda\right) T^\text{v}_{22} + \tau \dot T^\text{v}_{22} &= -\eta \dot\Lambda ,
	\end{aligned}
	\label{CreepEvol}
\end{equation}
where we have introduced $\Lambda = \ln \lambda$. The above steps correspond to Eq.~(18) and to the first line of Section~4.3 in \citet{farina22}.

We now move on to formulating the creep problem in differential form. For this purpose, we differentiate $T_{11}$ in time ($\tau\dot{T}_{11} = 0$), and substitute \eqref{CreepEvol} to get
\begin{equation}
	\left(3\eta + \tau \lambda\sigma^{\text{e}\prime} - \phi\tau \, ( 2 T_{11}^\text{v} + T^\text{v}_{22} )\right) \dot\Lambda = T_{11}-\sigma^\text{e} ,
	\label{Farina}
\end{equation}
where $\lambda = \text{e}^\Lambda$.
This equation is equivalent to the one leading to Eq.~(26) of \citet{farina22}. Therein, the coefficient of $\dot\Lambda$ is then required to be positive, which according to the authors of the discussed publication, is only possible with $\phi=0$. This conclusion is somewhat hasty given that the coefficient of $\phi$ in the left-hand side of \eqref{Farina} might not take arbitrary values, see Eq.~\eqref{CreepEvol}.

Let us examine this issue in more detail by considering the differential system \eqref{CreepEvol}-\eqref{Farina} as a whole. In terms of the dimensionless time coordinate $t/\tau$, our system becomes
\begin{equation}
	\left\lbrace
	\begin{aligned} 
		\dot \vartheta_{1}^\text{v} &= 2 (1 - \phi \vartheta_{1}^\text{v}) \dot\Lambda -  \vartheta_{1}^\text{v} ,\\
		\dot \vartheta^\text{v}_{2} &= (\phi \vartheta^\text{v}_{2} - 1) \dot\Lambda - \vartheta^\text{v}_{2} ,\\
		\dot\Lambda &= \frac{\vartheta_{1}^\text{v} - \vartheta^\text{v}_{2}}{3 + m(\Lambda) - \phi\, (2 \vartheta_{1}^\text{v} + \vartheta^\text{v}_{2})} ,
	\end{aligned}
	\right.
	\label{DynSyst}
\end{equation}
where the dimensionless stresses $\vartheta^\text{v}_i = \tau T^\text{v}_{ii}/\eta$ have been introduced, and $m(\Lambda) = \tau\lambda\sigma^{\text{e}\prime}/\eta$ is positive. Upon substitution of $\dot \Lambda$ from the last line into the first two ones, Eq.~\eqref{DynSyst} takes the form of an autonomous dynamical system $\dot{\bf Q} = {\bf F}({\bf Q})$ for the vector ${\bf Q} = [\vartheta_{1}^\text{v}, \vartheta_{2}^\text{v}, \Lambda]^\text{T}$ with a suitable function ${\bf F}$. Its only parameters are $\phi$ and the rescaled Mooney parameters $\tau C_i/\eta$.

Obviously, the unique equilibrium state of the dynamical system \eqref{DynSyst} is the vector $\bar{\bf Q} = [0, 0, \bar\Lambda]^\text{T}$, where $\bar\Lambda$ can be deduced from the expression \eqref{StressElast} of $\sigma^\text{e}$ evaluated at equilibrium. This result matches Eq.~(19) of \cite{farina22}. In the equilibrium state, the Jacobian matrix $\partial{\bf F}/\partial{\bf Q}$ of the above dynamical system has the eigenvalues $0$, $-1$, and $\frac{- \bar m}{3 + \bar m}$, where $\bar m = m(\bar \Lambda)$. Since none of these eigenvalues has a positive real part, we conclude that the equilibrium state is never asymptotically unstable, showing that physically admissible creep responses are possible. It is worth pointing out that these stability properties do not depend on the choice of objective derivative (parameter $\phi$).

Next, let us study the initial evolution of the stretch for arbitrary imposed stresses $T_{11}$. Until loading starts, the material is in the undeformed state, ${\bf Q} = {\bf 0}$. Eq.~\eqref{Farina} reduces to $(\eta + \tau \mu)\dot\Lambda = \frac13 T_{11}$ in this state, thus showing that the stretch begins to increase towards the new equilibrium for positive imposed stresses{\,---\,}the opposite evolution of $\Lambda$ is obtained for negative imposed stresses. Therefore, the creep response is both physically acceptable in the initial state and near equilibrium with this model, for all $\phi$.

Nevertheless, at first sight, nothing prevents the solution of the creep problem to blow up in finite time if $\phi = \pm 1$, see the denominator of \eqref{DynSyst}\textsubscript{c}. To carry out this analysis, one might prove that the total mechanical energy is non-negative, and that it decays monotonously. This process is equivalent to the local thermodynamic analysis conducted by \citet{morro20} in the case $\phi=0$ and by \citet{haupt02} in the case $\phi=\pm 1$, besides the neglect of incompressibility and of the strain energy therein, respectively. Given that the kinematic energy, the strain energy and the viscous energy are positive quantities, they must remain bounded as well. Hence, blow-up is impossible. Contrary to the claim in \citet{farina22}, no restriction of the material parameters is needed to enforce physically admissible behaviour for this model when $\phi = \pm 1$.

\medskip

The present analysis can be extended to the model found in \citet{filograna09} (`second model' considered in the discussed study\footnote{Note the potential typo in \citet{farina22} where the study by \citet{saccomandi21} is linked to the model \eqref{Filograna}.This model is referred to as ``a different approach'' proposed in \cite{filograna09} by \citet{saccomandi21}.}), which is based on the rate equation
\begin{equation}
	\bm{T}^\text{E} + \tau\, \mathscr{D}(\bm{T}^\text{E}) = \bm{T}^\text{e} + 2\eta\bm{D}
	\label{Filograna}
\end{equation}
for the extra stress $\bm{T}^\text{E} = \bm{T} + p\bm{I}$. For the simple tension problem with Mooney--Rivlin elastic stress described above, the uniaxial stress is still given by $T_{11} = \sigma^\text{E}$, where we have used similar notations as earlier. Here, we arrive at
\begin{equation}
	\left(3\eta - \phi\tau\, (2 T^\text{E}_{11} + T^\text{E}_{22}) \right) \dot\Lambda = T_{11} - \sigma^\text{e} ,
	\label{FarinaFilograna}
\end{equation}
which matches Eq.~(25)\textsubscript{a} of \cite{farina22}. Using Eq.~(17) therein, we obtain a differential system for ${\bf Q} = [\vartheta_{1}^\text{E}, \vartheta_{2}^\text{E}, \Lambda]^\text{T}$ where similar notations were used as for the previous model.

For this model, the unique equilibrium state is determined by the same value of $\bar\Lambda$ as previously. In terms of the dimensionless time coordinate $t/\tau$, the corresponding dynamical system is characterised by the eigenvalues $0$, $-1$ and ${\bar m}/\left({\phi\, (2 \bar \vartheta_{1}^\text{e} + \bar \vartheta_{2}^\text{e}) - 3}\right)$, where the coefficient of $\phi$ in the denominator can take arbitrary values, see Eq.~\eqref{StressComp}. Thus, the equilibrium state can potentially become asymptotically unstable unless the Jaumann rate is used. Finally, the conclusion drawn by \cite{farina22} on the basis of disputable arguments is recovered: physically admissible behaviour can potentially only be obtained for $\phi = 0$ with this model.

The above discussions do not affect the analysis of the model by \cite{zhou91} in the discussed study either (`first model' therein), for which
\begin{equation}
	\bm{T} + \tau\, \mathscr{D}(\bm{T}) = -p\bm{I} + \bm{T}^\text{e} + 2\eta\bm{D} .
	\label{Zhou}
\end{equation}
Indeed, in this case, the creep problem is described by a differential equation where the coefficient $3\eta - 2\phi\tau T_{11}$ of $\dot\Lambda$ is constant, see Eq.~(21) of \cite{farina22}. Since $T_{11}$ can take arbitrary values, this model is only admissible for $\phi = 0$, in general.

Based on a dynamic tensile problem, the discussed study highlights also the need for the additional restriction $\eta/\mu > \tau$ in relation with the models \eqref{Filograna}-\eqref{Zhou} to ensure that the undeformed state remains asymptotically stable. This property can be inferred directly from the infinitesimal strain limit of \eqref{Filograna}-\eqref{Zhou}. In fact, both constitutive laws correspond to the tensorial SLS model \eqref{SLS}-\eqref{QSLS} with $\tau_\varepsilon = \eta/\mu$ at small strains. Thus, the above restriction is nothing else but the classical requirement $\tau_\varepsilon > \tau$ required for such rheologies \citep{carcione15}.

%%%%%%%%%%%%% References biblio %%%%%%%%%%%%%%%
\addcontentsline{toc}{section}{References}

%%------------ MRC BIBLIO
%\section*{\refname}
%\addcontentsline{toc}{section}{References}
\bibliography{biblio}{}

%\appendix

\end{document}